\documentclass[aps,reprint,prd]{revtex4-1}

\usepackage{graphicx}
\usepackage{indentfirst}
\usepackage{amsfonts}
\usepackage{amssymb}
\usepackage{microtype}
\usepackage[reqno,intlimits]{amsmath}
\usepackage[polish,english]{babel}

\newtheorem{theorem}{Theorem}

\begin{document}

\title{Nonexistence of the final first integral in the Zipoy-Voorhees
space-time}

\author{Andrzej J.~Maciejewski}
\email{maciejka@astro.ia.uz.zgora.pl}
\affiliation{J.~Kepler Institute of Astronomy, University of Zielona G\'ora,
Licealna 9, PL-65--417 Zielona G\'ora, Poland.}

\author{Maria Przybylska}
\email{M.Przybylska@proton.if.uz.zgora.pl} 
\affiliation{Institute of Physics, University of Zielona G\'ora, Licealna 9,
65--417 Zielona G\'ora, Poland }

\author{Tomasz Stachowiak} \email{stachowiak@cft.edu.pl} 
\affiliation{Center for Theoretical Physics PAS, Al. Lotnikow 32/46, 02-668
Warsaw, Poland }

\begin{abstract}
    We show that the geodesic motion in the Zipoy-Voorhees space-time is not
    Liouville integrable, in that there does not exist an additional first
    integral meromorphic in the phase-space variables.
\end{abstract}

\maketitle

\section{Introduction}

The question of the integrability of the test particle motion in the Zipoy-Voorhees
metric has recently attracted some attention, with both numerical
\cite{Brink,Lukes} and
analytical investigations \cite{Kruglikov}. The authors of
\cite{Kruglikov} were able to exclude the existence of some polynomial
first integrals, but they argue that some weaker form of integrability might
take place taking into account the results of \cite{Brink}. On the other hand,
the results of \cite{Lukes} indicate chaotic behaviour of the
system, but the region where that happens is very small when compared to the
phase space dominated by invariant tori, and the integration was performed with
the Runge-Kutta method of the 5th order only. Since it is known
\cite{Busvelle,Yao}
that integrable systems can exhibit numerical chaos (particularly for the R-K
method), the results of \cite{Lukes} should be taken cautiously. Our own
numerical integration produced a Poincar\'e section visibly shifted from the
one in \cite{Lukes} (see the end of section {\bf IV}), and since we used a more
accurate method, it poses the question of whether the picture would be further
deformed as the precision was increased. In other words, to decide on
the integrability of the problem, a rigorous mathematical analysis is required
rather than numerical simulations.

The physical problem and its significance
are the same as in the classical paper by Carter \cite{Carter} -- that the
existence of an additional first integral in the Kerr space-time makes the
problem completely
integrable. Carter's integral is not generated by a Killing vector, so it is
not a usual symmetry of the manifold, but it is quadratic in momenta which
has important consequences. Such integrals translate into
the separability of the Hamilton-Jacobi equation and d'Alembertian
\cite{Waksjo}, which in turn appears in the Teukolsky \cite{Teukolsky}
equation. That is to say, both
the classical problem of particle motion in this space-time, the linear
perturbation equations governing the gravitational waves and potentially
quantum equations in that background become considerably
easier to solve. This fact is also used in numerical approaches, when
trying to determine possible spectra of gravitational radiation in anticipation
of the observed data \cite{Drasco}.

It is then natural to analyze other space-times which could serve as models of
compact objects, and the stationary axisymmetric ones are one direction to
explore. However, despite some numerical evidence \cite{Brink}  we find that the
particular Zipoy-Voorhees metric with the parameter $\delta=2$ is not
integrable. To be more precise, we consider the motion of a test particle as a
Hamiltonian system with $n$ degrees of freedom and ask for the existence of an additional
constant of motion $I_n$
that would yield {\em Liouvillian integrability} with respect to the canonical
Poisson bracket $\{\cdot\,,\cdot\}$. That is, for all first integrals $I_k$ we
would have
$\{I_k,I_l\}=0$, where the Hamiltonian is included as $H=I_1$, and
$I_2,\ldots,I_{n-1}$ are also already known. 
It turns out, that no such first integral can be found
in the class of meromorphic functions, and we will use the differential
Galois theory to prove that. Recall that a function is called meromorphic when
its singularities (if it has any) are just poles; so by allowing first integrals
that are potentially singular at some points of the phase space we are
considering a fairly wide class of functions.

The reason for using this particular theory is that it gives the strongest known 
necessary conditions for the integrability of dynamical systems. It was used
for proving the nonintegrability of the hardest problems of classical mechanics,
like the three-body problem \cite{Weil,Tsygvintsev},  which had been open for
centuries.  For an accessible overview of applications, see \cite{MorOver}.

\section{Formulation of the problem}
The Zipoy-Voorhees metric under consideration is given by
\begin{equation}
\begin{split}
    \mathrm{d}s^2 =&
    -\left(\frac{x-1}{x+1}\right)^2\mathrm{d}t^2+
    \frac{(x+1)^3(1-y^2)}{x-1}\mathrm{d}\phi^2\\&+
    \frac{(x^2-1)^2(x+1)^4}{(x^2-y^2)^3}\left(
    \frac{\mathrm{d}x^2}{x^2-1}+\frac{\mathrm{d}y^2}{1-y^2}\right),
\label{metric}
\end{split}
\end{equation}
where $x$, $y$ and $\phi$ form the prolate spheroidal coordinates.
Instead of working directly with the geodesic equations we take the Hamiltonian
approach with
\begin{equation}
\begin{split}
    H =& \frac12 g^{\alpha\beta}p_{\alpha}p_{\beta} =
    -\frac{(x+1)^2}{2(x-1)^2}p_0^2+
    \frac{(x^2-y^2)^3}{2(x-1)(x+1)^5}p_1^2 \\&+
    \frac{(x^2-y^2)^3(1-y^2)}{2(x-1)^2(x+1)^6}p_2^2 +
    \frac{x-1}{2(x+1)^3(1-y^2)}p_3^2,
    \label{Ham}
\end{split}
\end{equation}
where the canonical coordinates are
\begin{equation}
    q_0 = t,\; q_1 = x,\; q_2 = y,\; q_3 = \phi.
\end{equation}
The equations then read
\begin{equation}
    \left\{
\begin{aligned}
    \frac{dq_i}{d\tau} &= \frac{\partial H}{\partial p_i},\\
    \frac{dp_i}{d\tau} &= -\frac{\partial H}{\partial q_i},
    \label{sys}
\end{aligned} \right.
\end{equation}
with $i=0,1,2,3$, and the normalization of four-velocities gives the
value of the (conserved) Hamiltonian to be $H = -\frac12 \mu^2$. The new
time parameter is the rescaled proper time $\mu\tau = s$, which allows us to
include the zero geodesics for photons without introducing another affine
parameter but with simply $\mu = 0$.

Since the metric has two  Killing vector fields $\partial_t$ and
$\partial_{\phi}$, the two momenta $p_0$ and $p_3$ are conserved. Together with
the Hamiltonian they provide three first integrals. The question then is
whether there exists one more first integral that would make the system
Liouville integrable. To answer this question we employ the differential
Galois approach to integrability. More specifically, we use the main theorem
of the Morales-Ramis theory \cite{Morales}.
\begin{theorem}
\label{MR}
    If a complex Hamiltonian system is completely integrable with meromorphic
    first integrals, then the identity component of the differential Galois
    group of the variational and the normal variational equations along any
    nonconstant particular solution of this system is Abelian.
\end{theorem}

\section{Theoretical setting}

Let us try to explain the involved mathematics somewhat.  For detailed
exposition of the differential Galois theory the reader is referred to books
\cite{Kaplansky,Singer}. The  Morales-Ramis theory is exposed in
\cite{Morales, Audin}, and a short introduction with application to another
relativistic system can be found in \cite{scalarfields}.

To describe the differential Galois approach to the integrability we consider 
a general system of differential equations
\begin{equation}
\label{ds}
\frac{\mathrm{d} u}{\mathrm{d} \tau} = f(u), \qquad u=(u_1,\ldots, u_m).
\end{equation}
We assume that the right-hand sides 
\[ f(u)=(f_1(u), \ldots, f_m(u)),
\]
are meromorphic in the considered domain.  Let $\varphi(\tau)$ be a
nonequilibrium solution of this system. Then the variational equation  (VE)
along this solution have the form
\begin{equation}
\frac{\mathrm{d} \xi}{\mathrm{d} \tau} = A(\tau) \xi, \qquad  
A(\tau) = \dfrac{\partial f}{\partial u}(\varphi(\tau)).
    \label{VE}
\end{equation}
It is not difficult to prove that if the original system has an analytic first
integral $I(u)$, then the variational equation have a time-dependent  first
integral $I^\circ(\tau,\xi)$ which is polynomial in $\xi$. Similarly, one can
show that if $I(u)$ is a meromorphic first integral, then the variational
equations~\eqref{VE} have a first integral $I^\circ(\tau,\xi)$  which is
rational in $\xi$.  The Ziglin lemma, see p. 64 in \cite{Audin},  says that if
the system~\eqref{ds} has $1\leq k <m$ functionally independent first integrals
$I_j(u)$,  $j=1,\ldots, k$, then the variational equations ~\eqref{VE} have the
same number functionally independent first integrals $I_j^\circ(\tau,\xi)$
which are rational functions of $\xi$.

In the considered theory, time is assumed to be a complex variable,
and for complex $\tau\in\mathbb{C}$, the solution $\varphi(\tau)$  can have
singularities.  Assume that $\tau_0\in\mathbb{C}$ is not a singular point of
$\varphi(\tau)$. Then in a neighborhood of  $\tau_0$  there exist $m$
linearly independent solutions of the variational equations~\eqref{VE}. They are
the columns of the fundamental matrix $\Xi(\tau)$ of the system~\eqref{VE}.
This matrix can be analytically continued along an arbitrary path $\sigma$ on
the complex plane  avoiding the singularities of  the solution
$\varphi(\tau)$. Assume
that $\sigma$ is such a closed path, or loop, with the base point $\tau_0$.  Let
$\widehat{\Xi}(\tau)$ be a continuation of $\Xi(\tau)$. Solutions of a system
of $n$ linear equations form a linear $n$-dimensional space. Thus, in a
neighborhood of $\tau_0$, each column of  $\widehat{\Xi}(\tau)$  is a linear
combination of columns of $\Xi(\tau)$. We can write this fact in the form
$\widehat{\Xi}(\tau)= \Xi(\tau)M_\sigma$, where  $M_\sigma$ is a complex
nonsingular matrix, i.e., $M_\sigma\in\mathrm{GL}(m,\mathbb{C})$.  In fact, the
matrix $M_\sigma$ depends only on the homotopy class $[\sigma]$ of the loop.
Taking all loops with the base point $\tau_0$ we obtain an a group of
matrices $\mathcal{M}\subset  \mathrm{GL}(m,\mathbb{C})$ which is called the
monodromy  group of the equation~\eqref{VE}.

One can show that if $I^\circ(\tau, \xi)$ is a first integral of~\eqref{VE},
then  $I^\circ(\tau, \xi)=I^\circ(\tau, M\xi)$ for an arbitrary
$M\in\mathcal{M}$, and for an arbitrary $\tau$ from a neighborhood of
$\tau_0$. In other words, if the original system has a meromorphic first
integral, then the monodromy group has a rational invariant. Hence, if the
system possesses a  big number of first integrals, then  the monodromy group of
variational equations cannot be too big because it has a large number of
independent rational invariants. This observation can be transformed into an
effective tool if we restrict our attention to Hamiltonian systems and the
integrability in the Liouville sense (complete integrability).  The above facts
are the basic ideas of the elegant Ziglin theory
\cite{Ziglin:82::b,Ziglin:83::b}. The problem in applying this theory is
that  the monodromy group is known for a very limited number of equations.

At the end of the previous century the Ziglin theory found a nice
generalization.    It was developed by Baider, Churchill, Morales, Ramis, Rod,
Sim\'o and Singer, see \cite{Audin,Churchill:96::b,Morales} and references
therein. In the context of Hamiltonian systems it is called the Morales-Ramis
theory, and  in some sense, it is an algebraic version of the Ziglin theory. It
formulates the necessary conditions for the integrability in terms of the
differential Galois group $\mathcal{G}\subset \mathrm{GL}(m,\mathbb{C})$ of the
variational equations. It is known that it is a linear algebraic group and that it
contains the the monodromy group.  By definition it is a subgroup of
$\mathrm{GL}(n,\mathbb{C})$ which preserves all polynomial relations between
solutions of the considered linear system, see \cite{Beukers}; and for a wide class of
equations it is generated by $\mathcal{M}$. The differential Galois
group  can serve for a study of integrability problems on the same footing as
the monodromy group. Namely, first integrals of~\eqref{VE} give rational
invariants of $\mathcal{G}$.

If the considered system is Hamiltonian then necessarily $m=2n$,  and  groups
$\mathcal{M}$ and $\mathcal{G}$ are subgroups of the symplectic group
$\mathrm{Sp}(2n,\mathbb{C})$. It can also be shown that the differential Galois group
$\mathcal{G}$ is a Lie group. If the system is completely integrable with $n$
meromorphic first integrals, then $\mathcal{G}$ has $n$ commuting  rational
invariants. The key lemma, see p. 72 in \cite{Audin}, states that if the above
is the case, then the Lie algebra of  $\mathcal{G}$  is Abelian. This 
means exactly that the identity component of  $\mathcal{G}$  is Abelian.

Determination of the differential Galois group is a difficult task.
Fortunately, in the context
of integrability, we need to know only if its identity component is Abelian.
If it is not Abelian then the system is nonintegrable. If we find that a
subsystem  of  VE has a non-Abelian identity component of the differential Galois
group, then conclusions are the same. This is why, in practice, we always
try to distinguish  a subsystem of VE.   It is easy to notice  that
$\psi(t)=f(\varphi(t))$ is a solution of~\eqref{VE}. Using it we can reduce the
dimension of VE by one. If the system~\eqref{ds} is Hamiltonian,  then first we
restrict it to the energy level of the particular solution. In effect, in
Hamiltonian  context we can easily distinguish a subsystem of variational
equations of dimension $2(n-1)$, which are called the normal variational
equations (NVE).

The difficulty of investigation of the differential Galois group of NVE depends,
among other things,  on the form of its matrix of coefficients, and so also on the
functional form  of particular solution. Quite often, by an introduction  of a
new independent variable $z=z(\tau)$ we can transform NVE to a system with
rational coefficients
\begin{equation}
    \frac{\mathrm{d}\phantom{l}}{\mathrm{d} z}\xi = B(z) \xi 
    \qquad B(z)=[b_{i,j}(z)], \quad b_{i,j}(z)\in\mathbb{C}(z).
    \label{linsys}
\end{equation}

The set of rational functions $\mathbb{C}(z)$ is a field, and equipped with the
usual
differentiation it becomes a differential field. Solutions of a system with rational
coefficients are typically not rational. The smallest differential field
containing all solutions of~\eqref{linsys} is called the Picard-Vessiot
extension of  $\mathbb{C}(z)$. The differential Galois group $\mathcal{G}$
of~\eqref{linsys} tells us how complicated its solutions are, i.e., if the
equations are
solvable.  Here solvability means that all solutions  can be obtained from a
rational function by a finite number of integrations, exponentiation and
algebraic operations \cite{Kaplansky}. This category of functions, called
Liouvillian,  includes all elementary functions, as well as some
transcendental, such as the logarithm or elliptic integrals, and is commonly
referred to as ``closed-form'' or ``explicit'' solutions.  The following
classical result connects the structure of $\mathcal{G}$ with the form of the solutions.
\begin{theorem}
System \eqref{linsys} is solvable, i.e., all its solutions are Liouvillian, if and only
if  the the identity component of its differential Galois group is solvable,
\end{theorem}
The connection of this theorem with integrability is the following. If
it is possible to show that either NVE,  or a subsystem of NVE are not
solvable, then the identity component of their differential Galois group is not
solvable, so, in particular is not Abelian. Thus, by Theorem~\ref{MR}, the
system is not integrable. The question of whether a given system with rational coefficient
is solvable can be resolved completely for a system of two equations (or one
equation of second order). In this case  there is an effective algorithm by
Kovacic for finding the Liouvillian solutions \cite{Kovacic}.  This algorithm
gives a definite answer, and if Liouvillian solutions exist it provides their
analytical form. There exist a similar, almost complete algorithm for systems
of three equations and some partial results for systems of four equations.

\section{Proof of nonintegrability}

The plan of attack is thus to look for particular solutions for which the
NVE has a block structure so that a two-dimensional subsystem can be separated.
We then rewrite it as a second-order linear differential equation with rational
coefficients and apply the Kovacic algorithm to see if it has any Liouvillian
solutions. Note that the system has no external parameters, and only the values
of particular first integrals enter as internal parameters. They are synonymous
with initial conditions, so that if we manage to find just one solution, for
particular values of $\mu$, $p_0$ and $p_3$, such that the respective NVE is
unsolvable, we will have proven that there cannot exist another first integral
over the whole phase space.

It might so happen, unlike in the Carter case, that the system exhibits some
particular invariant
set on which there exists an additional integral. For example, one could have
$\dot{I_4}=H$, which would mean that $I_4$ is conserved 
on the zero-energy hypersurface $\mu=0$, which is clearly a physically
distinguished case. We will then have to look for particular solutions on those
sets to make the results even more restrictive than just the lack of
a global first integral.

The obvious particular solution to look at is a particle moving along a
straight line, through the center in the equatorial plane, which in prolate
coordinates means $y=0$ and $p_3=0$. The nontrivial equations then read:
\begin{equation}
\begin{aligned}
    \frac{\mathrm{d}t}{\mathrm{d}\tau} &= -\frac{(x+1)^2p_0}{(x-1)^2},\\
    \frac{\mathrm{d}x}{\mathrm{d}\tau} &= \frac{p_1 x^6}{(x-1)(x+1)^5},\\
    \frac{\mathrm{d}p_1}{\mathrm{d}\tau} &= 
    p_1^2 \frac{x^5(3-2x)}{(x+1)^6(x-1)^2} - p_0^2\frac{2(x+1)}{(x-1)^3}.
\end{aligned}
\end{equation}
Or, upon rescaling time by
\begin{equation}
    \mathrm{d}\tau = \frac{(x-1)^2(x+1)^3}{x^3}\mathrm{d}u,
\end{equation}
we have
\begin{equation}
\begin{aligned}
    \dot{x} &= \frac{p_1 x^3(x-1)}{(x+1)^2},\\
    \dot{p_1} &= p_1^2 \frac{x^2(3-2x)}{(x+1)^3}
    - p_0^2\frac{2(x+1)^4}{x^3(x-1)},
\end{aligned}
\end{equation}
where the dot denotes differentiation with respect to $u$, and we have omitted
the first equation, as the other two do not depend on $t$. This 
two-dimensional subsystem defines the particular solution around which we will
construct the NVE as mentioned before. The
conservation of the Hamiltonian now reads
\begin{equation}
    -\frac12\mu^2 = -\frac{p_0^2(x+1)^8+p_1^2x^6(1-x^2)}{2(x-1)^2(x+1)^6},
\end{equation}
which together with the equation for $\dot x$ yields
\begin{equation}
    \dot{x}^2 = (x^2-1)(p_0^2(x+1)^2-\mu^2(x-1)^2),
    \label{spec_sol}
\end{equation}
so that $x(u)$ is expressible by the Jacobi elliptic functions. This fact is
important, as we will change the independent variable from $u$ to $x$ which
is permissible (does not change the identity component of the Galois group)
only if the function $x(u)$ defines a finite cover of the
complex plane \cite{Singer}.

The variational equations along this solution separate so that the NVE read
\begin{equation}
\begin{split}
    \dot\xi_1 &= \frac{x^3}{(1+x)^3}\xi_2,\\
    \dot\xi_2 &= 3p_1^2\frac{x(x-1)}{(x+1)^2}\xi_1,
    \label{nve_u}
\end{split}
\end{equation}
where the variations $\xi$ correspond to the perturbations of variables $y$ and
$p_2$. This is another step of the reduction mentioned in the previous section
-- the particular solution only have $x$ and $p_1$ components, and the NVE only
has components in the orthogonal directions of $y$ and $p_2$.

Introducing a new dependent variable
\begin{equation}
    \xi =
    \frac{p_0^{1/2}x^{5/2}(x-1)^{1/4}}
    {(x+1)^{5/4}\left(p_0^2(x+1)^2-\mu^2(x-1)^2\right)^{1/4}}\xi_2,
\end{equation}
and taking $x$ as the new independent variable, the NVE can be brought to the
standard form of
\begin{equation}
    \xi''(x) = r(x)\xi(x),
    \label{normal}
\end{equation}
with the rational coefficient $r$
\begin{equation}
    r(x) := \frac{R(x)}{4x^2(x^2-1)^2(p_0^2(x+1)^2-\mu^2(x-1)^2)^2},
\end{equation}
where $R$ is the following polynomial
\begin{equation}
\begin{split}
R(x) = & p_0^4(34x^2-40x+3)(x+1)^4\\
    &-6p_0^2\mu^2(6x^2-10x+1)(x^2-1)^2\\
       &+\mu^4(22x^2-20x+3)(x-1)^4.
\end{split}
\end{equation}

Since for all physical particles we have $p_0\neq0$ all the others parameters
can be rescaled by it
\begin{equation}
\mu\rightarrow\mu/p_0,\quad p_3\rightarrow p_3/p_0,
\end{equation}
which we use in what follows. 

As is customary, we will use the same notation as
in Kovacic's paper, adhering exactly to the steps and cases of the algorithm
\cite{Kovacic}. We note
that a linear equation like \eqref{normal} has local solutions in some
neighborhood of a singularity $x_{\star}$ of $r(x)$, which take the form
\begin{equation}
    \xi = (x-x_{\star})^{\alpha}g(x-x_{\star}),
\end{equation}
where $g$ is analytic at zero, $g(0)\neq0$, and $\alpha$ is called the
characteristic
exponent. The algorithm checks if it is possible to construct a global
solution, which, in the simplest case, is of the form
\begin{equation}
    \xi = P e^{\int\omega\,\mathrm{d}x}
    \label{pexp}
\end{equation}
for a polynomial $P(x)$ and rational $\omega(x)$. The
degree of $P$ is then linked with the exponents and that provides preliminary
restrictions on the parameters' values and integrability. 

The application of the algorithm itself is straightforward, and the only
complication is that the singularities and exponents might
depend on parameters. Fortunately there are only several special values of
$\mu$ that influence the outcome, and we outline the general steps in the two
subsections below. For details, the reader is referred to \cite{Kovacic}, 
and another version of the algorithm, as applied to the dynamical system of
the Bianchi VIII cosmology, can be found in \cite{MaciejkaB}.

\subsection{General $r(x)$}
The poles of $r(x)$ are 
\begin{equation}
    \left\{-1,0,1,\frac{\mu-1}{\mu+1},\frac{\mu+1}{\mu-1}\right\},
\end{equation}
and for all of them to be different we must have $\mu^2\neq 1$
and $\mu\neq 0$. All are of order 2, and the order at
infinity is 4, so that we need to check all the cases of the algorithm.

In case 1, the characteristic exponents $\alpha_c^{\pm}$ of \eqref{normal} form
the following set
\begin{equation}
    \left\{ \left(0,1\right),\left(\tfrac94,-\tfrac54\right),
    \left(\tfrac32,-\tfrac12\right),
    \left(\tfrac34,\tfrac14\right),
    \left(\tfrac54,-\tfrac14\right),\left(\tfrac54,-\tfrac14\right)\right\},
\end{equation}
where the first pair corresponds to $\infty$,
and the combinations
\begin{equation}
    d = \alpha_{\infty}^{\pm} - \sum_{c,s}\alpha_c^s
\end{equation}
give only nine non-negative integers (not all distinct) as possible degrees of
the appropriate polynomial $P$, which enters into the solution of
\eqref{normal}. However, the respective test solutions of the form as in \eqref{pexp}
require that $\mu=0$, and
have to be discarded so that this case cannot hold.

In case 2, the families of exponents $E_c$ are
\begin{equation}
\begin{split}
    \left\{(0,2,4),(9,2,-5),(6,2,-2)\right.,\\
    \left.(3,2,1),(5,2,-1),(5,2,-1)\right\},
\end{split}
\end{equation}
which in turn give 131 possible integer degrees for the appropriate polynomial.
Checking them one by one, we find that they require $\mu=\pm 1$ in order to
form a solution, so that this case can be discarded as well under the current
assumptions.

In case 3, the families $E_c$ contain $6\times 13=78$ numbers, which make
4826809 combinations for $d$ out of which 230856 are non-negative integers. We
thus first resort to checking for the presence of logarithms in the solutions,
which would prevent this case \cite{Singer}.

The only poles with integer difference in the exponents are 0 and $\infty$.
Using the Frobenius method \cite{Whittaker}, we get the two independent
solutions around zero
\begin{equation}
\begin{split}
    v_1 =& x^{3/2}\left(1 +\frac{5(\mu^2-2)}{3(1-\mu^2)}x+
        \frac{23\mu^4-38\mu^2+65}{12(1-\mu^2)^2}x^2+
    \ldots\right),\\
    v_2 =&  x^{-1/2}\left(\frac19(\mu^2-1)+
    \frac59(\mu^2-2)x +\ldots\right)\\
    &+ (5-\mu^2)\ln(x)v_1.
\end{split}
\end{equation}
As can be seen, the logarithm is present when $\mu^2\neq 5$, 
and since the solutions around $\infty$ do not have logarithms at all, 
the only possibility for case 3 left here is with $\mu^2=5$.

\subsection{Special subcases}

In order to exclude the special energy hypersurfaces $\mu=0$, $\mu^2=1$ and
$\mu^2 = 5$, we have to resort to a
more general particular solution, namely one with $p_3\neq0$. As already
mentioned, it is enough to find one solution for each such surface, and that
means we can take a specific value of $p_3$. The corresponding NVE will only
have numeric coefficients, and checking for its Liouvillian solutions is much
easier, for it suffices to use one of available implemented routines, for
example the ``kovacicsols'' of the symbolic system Maple.

The solution will also be expressible by (hyper)elliptic function as defined
by the Hamiltonian constraint
\begin{equation}
    \dot{x}^2 = \frac{(x-1)(x+1)^5 - (x-1)^4p_3^2 - (x^2-1)^3-\mu^2}
    {(x+1)^2},
\end{equation}
and the counterparts of the NVE given in \eqref{nve_u} will read
\begin{equation}
\begin{split}
    \dot\xi_1 &= \frac{x^3}{(1+x)^3}\xi_2,\\
\dot\xi_2 &= \frac{(x-1)\left(3p_1^2x^4-(x^2-1)^2p_3^2\right)}
    {x^3(x+1)^2}\xi_1,
\end{split}
\end{equation}
We then proceed exactly as above, taking $x$ as the independent variable and
reducing the system to one equation of the form $\xi_2''=r\xi_2$. For each
hypersurface in question, the value of $p_3=1$ leads to NVE that are not
solvable with Liouvillian functions. 
This finishes the proof for all possible levels of the Hamiltonian.

\begin{figure*}[!ht]
\includegraphics[width=\textwidth]{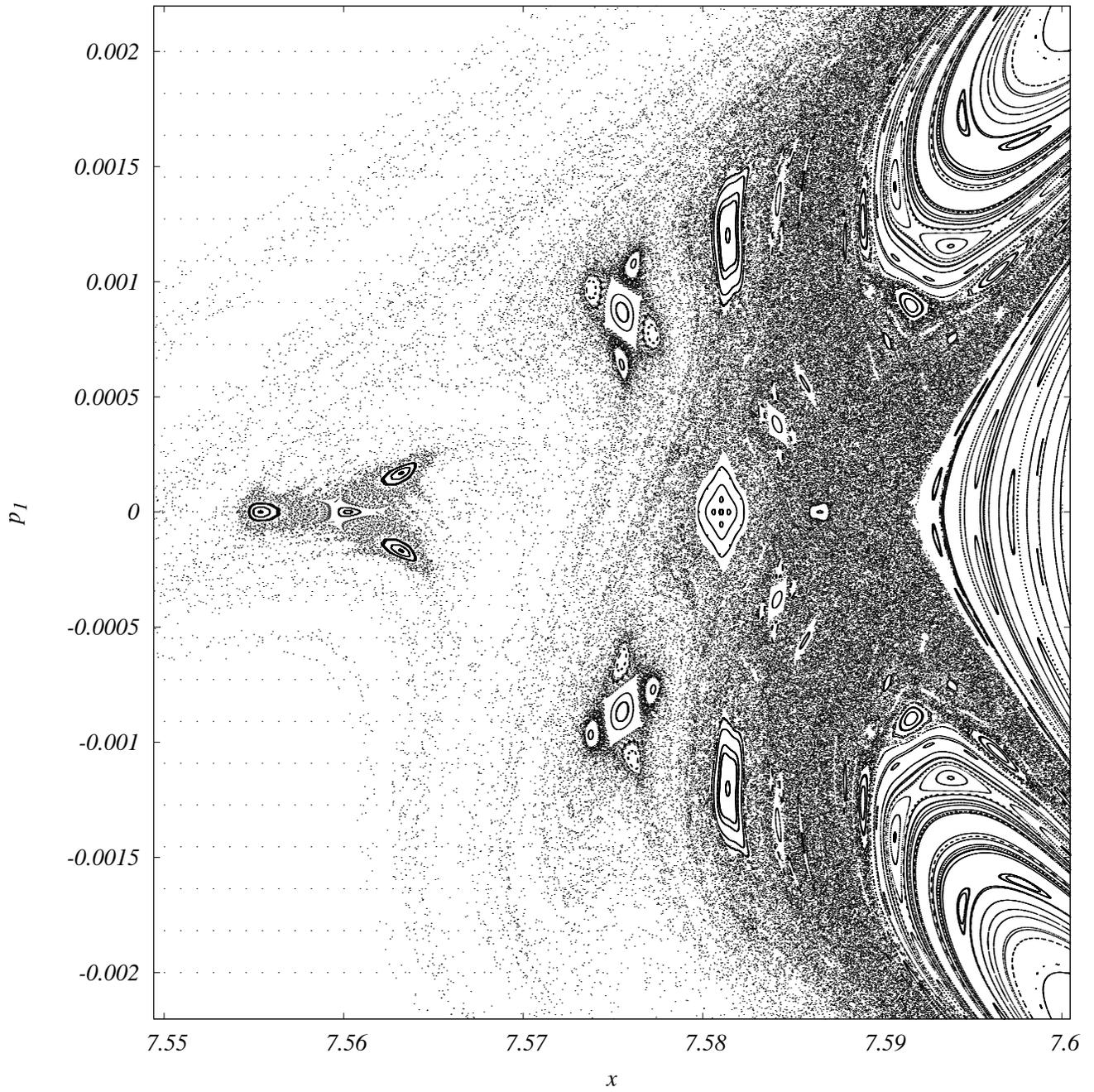}
\caption{Poincare section for the system \eqref{Ham} at $y=0$. The parameter
        values were: $p_0=0.95$, $p_3=3$, $\mu=1$.}
        \label{fig}
\end{figure*}

To further illustrate the complexity of this system, we have also obtained a
Poincar\'e section for the cross plane $y=0$ shown in Fig. \ref{fig}. The
numerical integrator was based on the Bulirsch-Stoer modified midpoint scheme
with Richardson extrapolation. We note that the special solution defined by
\eqref{spec_sol} lies entirely in the plane $y=0$ and is a trajectory beginning
and ending at the singularity so it does not contribute to the section. It also
lies outside the visible chaotic region, which is confined to a very small
subset of the phase space, as mentioned in \cite{Lukes}.

\section{General metric}

The above results carry, to some extent, to the general Zipoy-Voorhees metric
given by
\begin{equation}
\begin{split}
    \mathrm{d}s^2 &=
    -\left(\frac{x-1}{x+1}\right)^{\delta}\mathrm{d}t^2+
    \left(\frac{x+1}{x-1}\right)^{\delta}
    \Biggl((x^2-1)(1-y^2)\mathrm{d}\phi^2\Biggr.\\&+
    \Biggl.\left(\frac{x^2-1}{x^2-y^2}\right)^{\delta^2}(x^2-y^2)
    \left(
\frac{\mathrm{d}x^2}{x^2-1}+\frac{\mathrm{d}y^2}{1-y^2}\right)\Biggr),
\end{split}
\end{equation}
where $\delta\in\mathbb{R}$.
The main problem that arises for arbitrary $\delta$ is that the special
solution might no longer be a (hyper)elliptic function, because the
Hamiltonian now gives
\begin{equation}
\begin{split}
    \dot{x}^2 =& \frac{1}{x^2}\left(1-\frac{1}{x^2}\right)^{-\delta^2}
    (x+1)^{-2\delta}\Bigl((x+1)^{2\delta}(x^2-1)p_0^2\Bigr.\\
    &\Bigl.-(x-1)^{2\delta}p_3^2-(x^2-1)^{\delta+1}\mu^2\Bigr),
\end{split}
\end{equation}
so the right-hand side is not necessarily a polynomial or rational function.
Accordingly, the rationalization of the NVE might not preserve the identity
component of the differential Galois group.
However, when $\delta$ is rational we can still proceed by taking a new
dependent variable to be
\begin{equation}
    w:=\frac{x+1}{x-1},
\end{equation}
as this leads to the normal form \eqref{normal} which involves only integral powers
of $w$ and $w^{\delta}$. Assuming then that $\delta = p/q$, we can make the NVE
rational by taking $w^{1/q}$ as the new variable if need be. Unfortunately, the
number of poles (and their values) now depends on $p$ and $q$, so the Kovacic
algorithm has to be applied to each $\delta$ separately, but for each of them it
is as straightforward as above to use the Maple package, once suitable numeric
values of the parameters have been  chosen. For example, we have verified that
$\delta=1/2$ is also nonintegrable, confirming the numerical evidence of
\cite{Lukes} that for both $\delta > 1$ and $\delta < 1$, the general metric 
does not admit additional first integrals.

\section{Conclusions}

Our main result can be stated as 
\begin{theorem}
    There does not exist an additional, meromorphic first integral of the
    geodesic motion in the Zipoy-Voorhees metric \eqref{metric}, i.e., the
    system is not Liouville integrable.
\end{theorem}
This confirms the previous considerations of \cite{Kruglikov}
and goes much further than excluding first integrals polynomial in
momenta up to a certain fixed small degree. Meromorphic functions include not only
the analytic functions of both momenta and coordinates, but also rational
and transcendental ones as long as their singularities are just poles. In
particular, it follows that even if a conserved quantity exists, it cannot be
expressed by an explicit formula of the above type.
This result thus strongly reduces the possibility of using constants of motion
expansion in solving the equations of geodesic motion or gravitational waves
because the decomposition in terms of normal frequencies requires one to
calculate their values directly from the initial conditions of the coordinates
and momenta \cite{Drasco}. Of course, further techniques can be used to better
understand and describe the motion, especially in the region where the dynamics
is regular, but the fundamental physical property of this space-time is that no
additional conservation law holds.

\acknowledgements
This research has been supported by grant No.
DEC-2011/02/A/ST1/00208 of National Science Centre of Poland.

\end{document}